\documentclass[aps,prc,superscriptaddress,showpacs,floatfix,twocolumn]{revtex4}

\usepackage{graphicx}	% Include figure files

\begin{document}

\title{Measurement of Single Electron Event Anisotropy in Au+Au 
Collisions at $\sqrt{s_{NN}}$ = 200 GeV}

\newcommand{\abilene}{Abilene Christian University, Abilene, TX 79699, USA}
\newcommand{\acadsin}{Institute of Physics, Academia Sinica, Taipei 11529, Taiwan}
\newcommand{\banaras}{Department of Physics, Banaras Hindu University, Varanasi 221005, India}
\newcommand{\barc}{Bhabha Atomic Research Centre, Bombay 400 085, India}
\newcommand{\bnl}{Brookhaven National Laboratory, Upton, NY 11973-5000, USA}
\newcommand{\caucr}{University of California - Riverside, Riverside, CA 92521, USA}
\newcommand{\ciae}{China Institute of Atomic Energy (CIAE), Beijing, People's Republic of China}
\newcommand{\cns}{Center for Nuclear Study, Graduate School of Science, University of Tokyo, 7-3-1 Hongo, Bunkyo, Tokyo 113-0033, Japan}
\newcommand{\columbia}{Columbia University, New York, NY 10027 and Nevis Laboratories, Irvington, NY 10533, USA}
\newcommand{\dapnia}{Dapnia, CEA Saclay, F-91191, Gif-sur-Yvette, France}
\newcommand{\debrecen}{Debrecen University, H-4010 Debrecen, Egyetem t{\'e}r 1, Hungary}
\newcommand{\fsu}{Florida State University, Tallahassee, FL 32306, USA}
\newcommand{\gsu}{Georgia State University, Atlanta, GA 30303, USA}
\newcommand{\hiroshima}{Hiroshima University, Kagamiyama, Higashi-Hiroshima 739-8526, Japan}
\newcommand{\ihepprot}{Institute for High Energy Physics (IHEP), Protvino, Russia}
\newcommand{\isu}{Iowa State University, Ames, IA 50011, USA}
\newcommand{\jinrdubna}{Joint Institute for Nuclear Research, 141980 Dubna, Moscow Region, Russia}
\newcommand{\kaeri}{KAERI, Cyclotron Application Laboratory, Seoul, South Korea}
\newcommand{\kangnung}{Kangnung National University, Kangnung 210-702, South Korea}
\newcommand{\kek}{KEK, High Energy Accelerator Research Organization, Tsukuba-shi, Ibaraki-ken 305-0801, Japan}
\newcommand{\kfki}{KFKI Research Institute for Particle and Nuclear Physics (RMKI), H-1525 Budapest 114, POBox 49, Hungary}
\newcommand{\korea}{Korea University, Seoul, 136-701, Korea}
\newcommand{\kurchatov}{Russian Research Center ``Kurchatov Institute", Moscow, Russia}
\newcommand{\kyoto}{Kyoto University, Kyoto 606-8502, Japan}
\newcommand{\labllr}{Laboratoire Leprince-Ringuet, Ecole Polytechnique, CNRS-IN2P3, Route de Saclay, F-91128, Palaiseau, France}
\newcommand{\lawllnl}{Lawrence Livermore National Laboratory, Livermore, CA 94550, USA}
\newcommand{\losalamos}{Los Alamos National Laboratory, Los Alamos, NM 87545, USA}
\newcommand{\lpc}{LPC, Universit{\'e} Blaise Pascal, CNRS-IN2P3, Clermont-Fd, 63177 Aubiere Cedex, France}
\newcommand{\lund}{Department of Physics, Lund University, Box 118, SE-221 00 Lund, Sweden}
\newcommand{\muenster}{Institut f\"ur Kernphysik, University of Muenster, D-48149 Muenster, Germany}
\newcommand{\myongji}{Myongji University, Yongin, Kyonggido 449-728, Korea}
\newcommand{\nagasaki}{Nagasaki Institute of Applied Science, Nagasaki-shi, Nagasaki 851-0193, Japan}
\newcommand{\newmex}{University of New Mexico, Albuquerque, NM 87131, USA}
\newcommand{\nmsu}{New Mexico State University, Las Cruces, NM 88003, USA}
\newcommand{\ornl}{Oak Ridge National Laboratory, Oak Ridge, TN 37831, USA}
\newcommand{\orsay}{IPN-Orsay, Universite Paris Sud, CNRS-IN2P3, BP1, F-91406, Orsay, France}
\newcommand{\pnpi}{PNPI, Petersburg Nuclear Physics Institute, Gatchina, Russia}
\newcommand{\riken}{RIKEN (The Institute of Physical and Chemical Research), Wako, Saitama 351-0198, JAPAN}
\newcommand{\rikjrbrc}{RIKEN BNL Research Center, Brookhaven National Laboratory, Upton, NY 11973-5000, USA}
\newcommand{\saispbstu}{St. Petersburg State Technical University, St. Petersburg, Russia}
\newcommand{\saopaulo}{Universidade de S{\~a}o Paulo, Instituto de F\'{\i}sica, Caixa Postal 66318, S{\~a}o Paulo CEP05315-970, Brazil}
\newcommand{\seoulnat}{System Electronics Laboratory, Seoul National University, Seoul, South Korea}
\newcommand{\stonybrkc}{Chemistry Department, Stony Brook University, SUNY, Stony Brook, NY 11794-3400, USA}
\newcommand{\stonycrkp}{Department of Physics and Astronomy, Stony Brook University, SUNY, Stony Brook, NY 11794, USA}
\newcommand{\subatech}{SUBATECH (Ecole des Mines de Nantes, CNRS-IN2P3, Universit{\'e} de Nantes) BP 20722 - 44307, Nantes, France}
\newcommand{\tenn}{University of Tennessee, Knoxville, TN 37996, USA}
\newcommand{\titech}{Department of Physics, Tokyo Institute of Technology, Tokyo, 152-8551, Japan}
\newcommand{\tsukuba}{Institute of Physics, University of Tsukuba, Tsukuba, Ibaraki 305, Japan}
\newcommand{\vandy}{Vanderbilt University, Nashville, TN 37235, USA}
\newcommand{\waseda}{Waseda University, Advanced Research Institute for Science and Engineering, 17 Kikui-cho, Shinjuku-ku, Tokyo 162-0044, Japan}
\newcommand{\weizmann}{Weizmann Institute, Rehovot 76100, Israel}
\newcommand{\yonsei}{Yonsei University, IPAP, Seoul 120-749, Korea}
\affiliation{\abilene}
\affiliation{\acadsin}
\affiliation{\banaras}
\affiliation{\barc}
\affiliation{\bnl}
\affiliation{\caucr}
\affiliation{\ciae}
\affiliation{\cns}
\affiliation{\columbia}
\affiliation{\dapnia}
\affiliation{\debrecen}
\affiliation{\fsu}
\affiliation{\gsu}
\affiliation{\hiroshima}
\affiliation{\ihepprot}
\affiliation{\isu}
\affiliation{\jinrdubna}
\affiliation{\kaeri}
\affiliation{\kangnung}
\affiliation{\kek}
\affiliation{\kfki}
\affiliation{\korea}
\affiliation{\kurchatov}
\affiliation{\kyoto}
\affiliation{\labllr}
\affiliation{\lawllnl}
\affiliation{\losalamos}
\affiliation{\lpc}
\affiliation{\lund}
\affiliation{\muenster}
\affiliation{\myongji}
\affiliation{\nagasaki}
\affiliation{\newmex}
\affiliation{\nmsu}
\affiliation{\ornl}
\affiliation{\orsay}
\affiliation{\pnpi}
\affiliation{\riken}
\affiliation{\rikjrbrc}
\affiliation{\saispbstu}
\affiliation{\saopaulo}
\affiliation{\seoulnat}
\affiliation{\stonybrkc}
\affiliation{\stonycrkp}
\affiliation{\subatech}
\affiliation{\tenn}
\affiliation{\titech}
\affiliation{\tsukuba}
\affiliation{\vandy}
\affiliation{\waseda}
\affiliation{\weizmann}
\affiliation{\yonsei}
\author{S.S.~Adler}	\affiliation{\bnl}
\author{S.~Afanasiev}	\affiliation{\jinrdubna}
\author{C.~Aidala}	\affiliation{\bnl}
\author{N.N.~Ajitanand}	\affiliation{\stonybrkc}
\author{Y.~Akiba}	\affiliation{\kek} \affiliation{\riken}
\author{J.~Alexander}	\affiliation{\stonybrkc}
\author{R.~Amirikas}	\affiliation{\fsu}
\author{L.~Aphecetche}	\affiliation{\subatech}
\author{S.H.~Aronson}	\affiliation{\bnl}
\author{R.~Averbeck}	\affiliation{\stonycrkp}
% ...
\collaboration{PHENIX Collaboration} \noaffiliation

\date{\today}

\begin{abstract}

The transverse momentum dependence of the azimuthal anisotropy parameter $v_{2}$,
the second harmonic of the azimuthal distribution, for electrons 
at mid-rapidity ($|\eta|<0.35$) has been measured with the PHENIX detector in
Au+Au collisions at $\sqrt{s_{NN}}$ = 200 GeV. 
The measurement was made with respect to the reaction plane defined at high rapidities ($|\eta|=3.1-3.9$).
From the result we have measured the $v_{2}$ of electrons from heavy flavor
decay after subtraction of the $v_{2}$ of electrons from other sources such as
photon conversions and Dalitz decay from light neutral mesons.
We observe a non-zero single electron $v_{2}$ with a 90 $\%$ confidence level in the intermediate $p_{T}$ region.

\end{abstract}

\maketitle

\section{Introduction}

The azimuthal anisotropy of particle emission is a powerful tool to study 
the early stage of ultra-relativistic nuclear collisions.
The spatial anisotropy in the initial stage of non-central 
nucleus-nucleus collisions is transferred into momentum anisotropy in the final state.
The azimuthal anisotropy is defined by
\begin{equation}
\frac{dN}{d\phi}=N_{0}\left\{1+\sum_{n}2v_{n}\cos(n(\phi-\Psi_{R.P.}))\right\},
\label{eq:EQ1}
\end{equation}
where  $N_{0}$ is a normalization constant, $\phi$ is the azimuthal angle of particles, 
and $\Psi_{R.P.}$ is the direction of the nuclear impact parameter ("reaction plane") in a given collision.
The harmonic coefficients, $v_{n}$, 
indicate the strength of the $n^{th}$ anisotropy.
The azimuthal anisotropy parameter $v_{2}$ 
(the second harmonic coefficient of the Fourier expansion of 
the azimuthal distribution) may be especially sensitive to the early pressure \cite{flowth0}. 
The transverse momentum ($p_{T}$) dependence of $v_{2}$ has been measured for identified particles 
at RHIC \cite{pidflowPHENIX, flowSTAR0,flowSTAR1, flowSTAR2}.
Previous measurements are limited to hadrons made of light quarks. 
These results show a clear mass dependence of $v_{2}$, which is well reproduced by a hydrodynamical calculation \cite{hydro} in the low 
$p_{T}$ region ($p_{T}<$ 2 GeV/$c$). 
The agreement is considered as evidence that the collective motion develops in the very early stages of the reaction. 
It is also observed that $v_{2}$ as a function of $p_{T}$ scales via the coalescence prescription, that is, $v_{2}/n$ as a function of $p_{T}/n$ is universal, where $n$ is 
the number of valence quarks plus valence anti-quarks.
This scaling behavior is consistent with the prediction of the quark coalescence model, which assumes a finite $v_{2}$ of quarks \cite{qcoales}. 
This suggests that the $v_{2}$ already develops in the partonic phase for hadrons made of light quarks.
In addition, if the $v_{2}$ of heavy quarks is non-zero,
it would support partonic level thermalization and  very high density at the early stage of the collisions.

Electrons are a useful tool to study the production of heavy quarks such as charm quarks.
In the PHENIX experiment at RHIC, transverse momentum spectra of single electrons have 
been measured in Au+Au collisions at $\sqrt{s_{NN}}$ = 130 GeV \cite{single} and 200 GeV \cite{esingle200}.
The results are consistent with that expected from semileptonic charm decays in
addition to decays of light mesons and photon conversions \cite{single}.
On the other hand, electrons originating from semileptonic decays of $D$ mesons have a significant angular 
deviation from the original $D$ meson direction.
The effect for $v_{2}$ has been studied in \cite{rapD} and \cite{resonance_decay}.
The results suggest that the effect is not significant for the decay electron $v_{2}$,
and the electron $v_{2}$ reflects the $v_{2}$ of $D$ meson.
Therefore the single electron $v_{2}$ measurement is a useful method for studying open charm $v_{2}$.

Currently the single electron spectra from PHENIX are consistent with two opposing scenarios:
(1) initial perturbative QCD charm production without final state interactions
and (2) complete thermal equilibrium for charmed hadrons \cite{thcharmsingle}.
Therefore the measurement of the azimuthal anisotropy of 
electrons from semileptonic charm decays could give us important new information regarding the 
charm dynamics in high-energy heavy-ion collisions.
The measurement is also important for the
quark coalescence model due to the large difference between the charm quark and 
light quark masses, and for the prediction of $v_{2}$ for the $J/\psi$ and the $D$ meson,
which contain charm quarks.

In this paper, we present the first measurement of the single electron $v_{2}$, which is expected to reflect 
the heavy flavor azimuthal anisotropy, below 4 GeV/$c$ with respect to the reaction plane in Au+Au 
collisions at $\sqrt{s_{NN}}$ = 200 GeV.
The single electron $v_{2}$ was measured by subtracting from the inclusive electron $v_{2}$ the $v_{2}$ of electrons from other sources
such as photon conversions and Dalitz decays from light neutral mesons.

\begin{figure}[htb]
\begin{center}
\includegraphics[width=1.0\linewidth,keepaspectratio,clip]{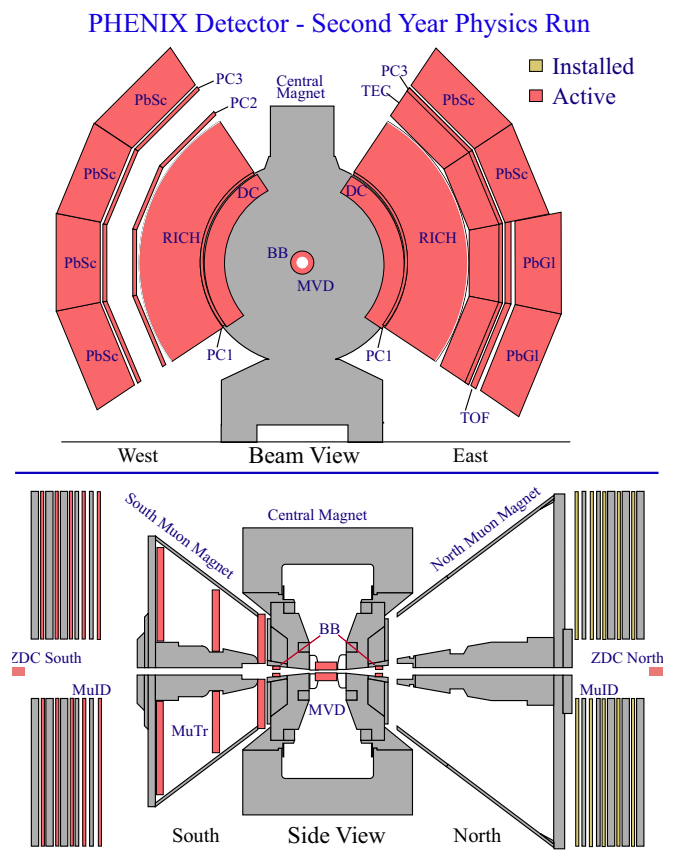}
\end{center}
\caption{\label{fig:phenixdetector}
PHENIX experiment configuration in Run2.
Top: Cross section perpendicular to the beam pipe. Bottom: East side view of the cross section 
along the beam pipe.}
\end{figure}

%-------\section{Data analysis}
\section{Data analysis}

About 16M minimum bias events in RHIC-Run2 (2001) for $\sqrt{s_{NN}}$ = 200 GeV are used in this analysis after a vertex cut is applied ($|z_{\rm vertex}|<$ 20 cm).
In this section
we present a brief overview of the PHENIX detectors \cite{PHENIXOverView} used in this analysis
and then present details of event selection, electron identification
and reaction plane determination.

\subsection{Overview of PHENIX detector}

PHENIX consists of four spectrometer arms (central arms and muon arms) and a set of global detectors.
The central arms are located east and west of the interaction region at
mid-rapidity.
The muon arms are located to the north and south at forward rapidity.
Fig.~\ref{fig:phenixdetector} shows the configuration of the central arms in Run2.
 
\begin{figure}[t]
\begin{center}
\includegraphics[width=0.7\linewidth]{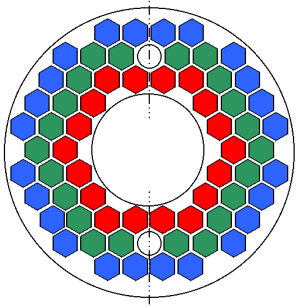}
\caption{\label{fig:bbcarray}
The configuration of the 64 PMTs of each BBC is shown.
From the hit position of the particles in each PMT, the azimuthal angle $\phi$ is calculated.}
\end{center}
\end{figure}

The global detectors consist of the Beam-Beam-Counters (BBCs) and the Zero-Degree Calorimeters (ZDCs).
These detectors provide the time of the Au+Au collision, the collision vertex, the event trigger and the collision centrality.
In this analysis the BBCs are also used to determine the reaction plane.
The BBCs are installed on the North and South sides of the collision point along the beam axis.
Each BBC is placed 144 cm from the center of the interaction region and surrounds the beam pipe.
This corresponds to a pseudorapidity range from 3.1 to 3.9 over the full azimuth.
Each BBC is composed of 64 elements; the configuration is shown in 
Fig.~\ref{fig:bbcarray}.
A single BBC element consists of a one-inch diameter mesh dynode photomultiplier tube mounted on a 3 cm long quartz radiator. 
The ZDC is a hadron calorimeter and measures the energy of spectator neutrons.
The ZDCs are located 18 m downstream and upstream along the beam axis, and 
each ZDC covers 2 mrad of forward angular cone, corresponding to $\eta>$6.0. 

The central arms are designed to track particles emitted from collisions, identify charged particles
and reconstruct invariant masses.
The central arms each cover the pseudorapidity range $|\eta| <$ 0.35 and $90^{\circ}$ in azimuthal angle.
The central arms consist of several subsystems.
In this analysis drift chambers (DCs), pad chambers (PCs), ring imaging Cherenkov counters (RICHs) 
and electromagnetic calorimeters (EMCals) are used.
The DCs are located between 2.0 and 2.4 m from the beam axis on each central arm and measure charged particle 
trajectories in the r-$\phi$ plane.
The central arms have three layers of PCs, which are multi-wire proportional chambers. 
The PCs are located at 2.4 m (PC1), 4.2 m (PC2) and 5.0 m (PC3) from the beam axis.
PC1 and PC3 are installed in each central arm, but PC2 is installed only in the west arm.
The PC measures 3-D space points along the straight line particle trajectories.
A RICH, the primary detector for electron identification, is installed in each central arm.
The RICH consists of a gas vessel, a thin reflector and a photon detector consisting of an array of PMTs.
During Run2 CO$_{2}$ was used as the Cherenkov radiator so
only pions with $p>4.7$ GeV/$c$ emit Cherenkov light in the RICH.
The EMCal is used to measure the spatial position and energy of electrons and photons. 
It covers the full central arm acceptance of $70^{\circ} < \theta < 110^{\circ}$ with each of the two walls
subtending $90^{\circ}$ in azimuth. 
One wall is comprised of four sectors of Pb-scintillator sampling calorimeter, and the other has two sectors of
Pb-scintillator and two of Pb-glass Cherenkov calorimeter.
The Central Magnet (CM) provides a magnetic field around the interaction vertex that is parallel to the beam. 
The CM allows momentum analysis of charged particles in polar angle range from $70^{\circ} < \theta < 110^{\circ}$
and provides a field integral of about 0.8 Tesla-meters \cite{PHENIXmagnet}.

\subsection{Event selection}
The event selection was done with the BBC and the ZDC in this analysis.
The minimum bias trigger requires a coincidence between 
north and south BBC signals.
The trigger included $92.2^{+2.5}_{-3.0}\%$ of the 6.9 barn Au+Au inelastic cross section \cite{minimumb}.
The event centrality is determined by combining information on spectator neutrons measured by the ZDC
and the charge sum information measured by the BBC.
The collision vertex point along the beam line is determined by the timing difference of the two BBCs.
We required $|z_{\rm vertex}|<$ 20 cm for this analysis.

\subsection{Charged particle selection and electron identification}
\label{sec-chpart}

\begin{figure}[tbh]
\begin{center}
\includegraphics[width=1.0\linewidth,clip]{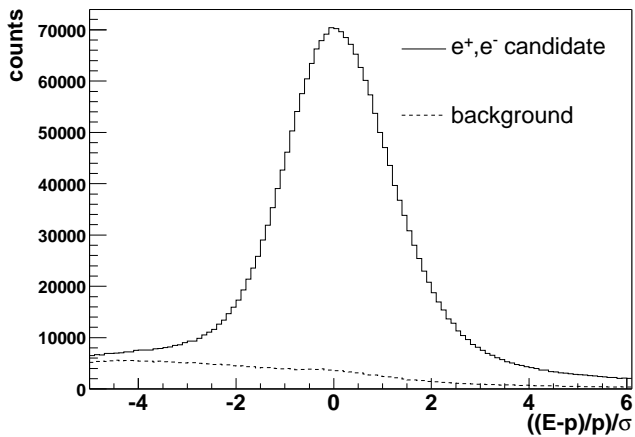}
\caption{\label{fig:eID}
$(E-p)/p/\sigma$ distribution.  
We require $-2 \sigma <(E-p)/p < 3 \sigma$ to reduce background from hadrons and photon conversions far from the vertex.
A background of less than 10 $\%$, caused by accidental association of tracks with RICH hits, remains.}
\end{center}
\end{figure}

Charged particle tracks are reconstructed by the DC
and the first pad chamber plane (PC1) installed in each central arm
together with the collision vertex determined by the BBC \cite{PHENIXdetectro1}.
In order for a reconstructed track to be selected, the track projection to the EMCal and the position of the associated hit in the EMCal must match within 2 standard deviations.
The electron candidates are required to have at least three associated hits in the RICH 
that pass a ring shape cut and are also required to pass a timing cut.
To reduce background from hadrons and photon conversions far from the vertex,
energy is measured in the EMCal, and momentum matching ($E/p$) is required.
Electrons deposit all of their energy in the EMCal; therefore the $E/p$ is approximately 1.0. 
In this analysis we require $-2 \sigma < (E-p)/p < 3 \sigma$ to reduce background.
Fig.~\ref{fig:eID} shows the $(E-p)/p/\sigma$ distribution.
Here the $\sigma$ means a standard deviation of $(E-p)/p$.
A background of less than 10 $\%$ remains, caused by accidental association of tracks with RICH hits. 
The background level is estimated by an event mixing method and is subtracted when we calculate the electron $v_{2}$.

\subsection{Reaction plane determination}

In this analysis the values of $v_{2}$ are calculated by the reaction plane method, which 
measures the azimuthal angle of the particle emission with respect to the reaction plane \cite{rpmethod}.
The azimuthal angle of the reaction plane for the $n^{th}$ harmonic is determined by \cite{rpmethod}
\begin{equation}
\psi^{\rm meas.}_{n}=\left(\tan^{-1}\frac{\sum_{i} w_{i}\sin(n\phi_{i})}{\sum_{i} w_{i} \cos(n\phi_{i})}\right)/n,
\label{eq:EQ2}
\end{equation}
where $\phi_{i}$ is the azimuthal angle of each particle used in the reaction plane determination and 
$w_{i}$ is the corresponding weight.
The azimuthal angle distribution of the particle emission measured with respect to the reaction plane 
can be written as Eq.~\ref{eq:EQ1}.
Due to finite reaction plane resolution, coefficients in the Fourier expansion of the azimuthal distribution 
with respect to the ``measured'' reaction plane ($v_{n}^{\rm meas.}$) are smaller than 
coefficients measured with respect to the ``real'' reaction plane ($v_{n}$).
The resolution correction necessary for $v^{\rm meas.}_{n}$ is given by;
\begin{equation}
v_{n}=v_{n}^{\rm meas.}/\sigma_{v_n}  
\label{eq:EQ3}
\end{equation}
where $v_{n}$ is the real coefficient and $\sigma_{v_n}$ is the reaction plane resolution for the $n^{th}$ harmonic.
The $\sigma_{v_n}$ is estimated with data using a formula shown in Ref.~\cite{rpmethod}.
The value of $v_{n}^{\rm meas.}$ is obtained by fitting the azimuthal distribution (relative to the reaction plane) with
\begin{equation}
\frac{dN}{d\phi}=N_{0}(1+2v_{n}^{\rm meas.}\cos(n\phi)),
\label{eq:EQ4}
\end{equation}
where $N_{0}$ and $v_{n}^{\rm meas.}$ are fitting parameters.  We can also calculate $v_{n}^{\rm meas.}$ directly by
\begin{equation}
v_{n}^{\rm meas.}=\langle\cos(n\phi)\rangle.
\label{eq:EQ5}
\end{equation}

In this analysis the $v_{2}$ is estimated by using the reaction plane found from the second harmonic ($n=2$),
since better accuracy of $v_{n}$ is obtained by 
using the same harmonic's reaction plane \cite{rpmethod}. 
The reaction planes are determined by using both BBCs.
In the PHENIX experiment the reaction plane is also determined by using the central arm detectors.
One of the key issues of the reaction plane determination is non-flow effects such as jets, resonance decays and HBT.
Since each BBC is roughly three units of pseudorapidity away from the central arms, 
it is expected that the non-flow effects are smaller there than in the central arm detectors \cite{pidflowPHENIX}.

\begin{figure}[tbh]
\begin{center}
\includegraphics[width=1.0\linewidth,clip]{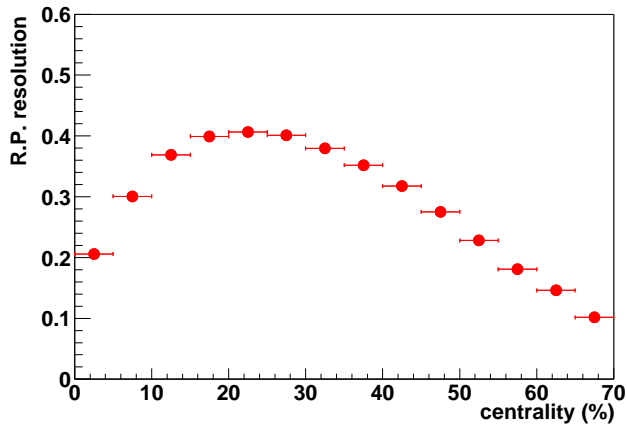}
\caption{\label{fig:bbcrpreso}
Centrality dependence of the combined reaction plane resolution that is determined by the BBC.}
\end{center}
\end{figure}

Using the BBC information the reaction plane is measured by
\begin{equation}
\psi=\left(\tan^{-1}\frac{\sum_{i=1}^{64} q_{i}\sin(2\phi_{i})}{\sum_{i=1}^{64} q_{i} \cos(2\phi_{i})}\right)/2
\label{eq:EQ6}
\end{equation}
where $\phi_{i}$ is the azimuthal angle of each PMT and $q_{i}$ is the charge information of each PMT.
Due to the random distribution of the impact parameter direction in collisions, 
the reaction plane should have an isotropic azimuthal distribution.
Because of the possible azimuthal asymmetries in the BBC response, however, the measured reaction plane distribution is anisotropic.
In this paper, we use the following two step method to correct the reaction plane.
First the distribution of $\sum_{i=1}^{64} q_{i}\sin(2\phi_{i})$ and $\sum_{i=1}^{64} q_{i} \cos(2\phi_{i})$
are recentered by subtracting $\langle \sum_{i=1}^{64} q_{i}\sin(2\phi_{i}) \rangle$ and 
$\langle \sum_{i=1}^{64} q_{i} \cos(2\phi_{i}) \rangle$ over all events ~\cite{rpmethod}:

\begin{equation}
\psi=\left(\tan^{-1} \frac{\sum_{i=1}^{64} q_{i}\sin(2\phi_{i})-\langle\sum_{i=1}^{64} q_{i}\sin(2\phi_{i})\rangle}
{\sum_{i=1}^{64} q_{i} \cos(2\phi_{i})-\langle\sum_{i=1}^{64} q_{i} \cos(2\phi_{i})\rangle}\right)/2.
\label{eq:EQ7}
\end{equation}
This method does not remove higher harmonic components of the determined reaction plane,
so we apply an additional correction method ~\cite{flattening}.
In this method flattening the reaction plane is accomplished by using a shift
\begin{equation}
n\psi^{\rm flat}=n\psi_{\rm obs.}+\Delta\psi,
\label{eq:EQ8}
\end{equation}
where $\Delta\psi$ is the correction factor for the reaction plane.
$\Delta\psi$ is determined by
\begin{equation}
\Delta\psi=\sum_{n} A_{n} \cos(2n\psi_{\rm obs.})+B_{n}\sin(2n\psi_{\rm obs.}). 
\label{eq:EQ9}
\end{equation}
$A_{n}$ and $B_{n}$ are defined by requiring the $n^{th}$ Fourier moment of 
the new reaction plane ($\psi^{\rm flat}$) to vanish.
\begin{eqnarray}
A_{n}&=&-\frac{2}{n}\langle\sin(2n\psi_{\rm obs.})\rangle \\
B_{n}&=&\frac{2}{n}\langle\cos(2n\psi_{\rm obs.})\rangle.
\label{eq:EQ10}
\end{eqnarray}
Since the reaction plane depends on collision centrality and z vertex, the
reaction planes are divided into 40 samples (20 centrality bins and 2 vertex bins),
and these corrections are determined independently for each sample.
A combined reaction plane, which is defined by weighted averaging the reaction plane angles 
obtained by the south side BBC and the north side BBC, is used to measure the $v_{2}$ 
in this analysis.   
The resolution of the combined reaction plane is estimated by using Eq. 11 in ~\cite{rpmethod}.
Figure ~\ref{fig:bbcrpreso} shows the centrality dependence of the resolution ($\sigma_{v_2}$).

\subsection{Inclusive electron $v_{2}$}
\label{incv2}
\section{Results and Discussion}

In this section we present a method to calculate the single electron $v_{2}$ from inclusive electrons and 
show its transverse momentum dependence.
\begin{figure*}[bth]
\begin{center}
\includegraphics[width=1.0\linewidth,clip]{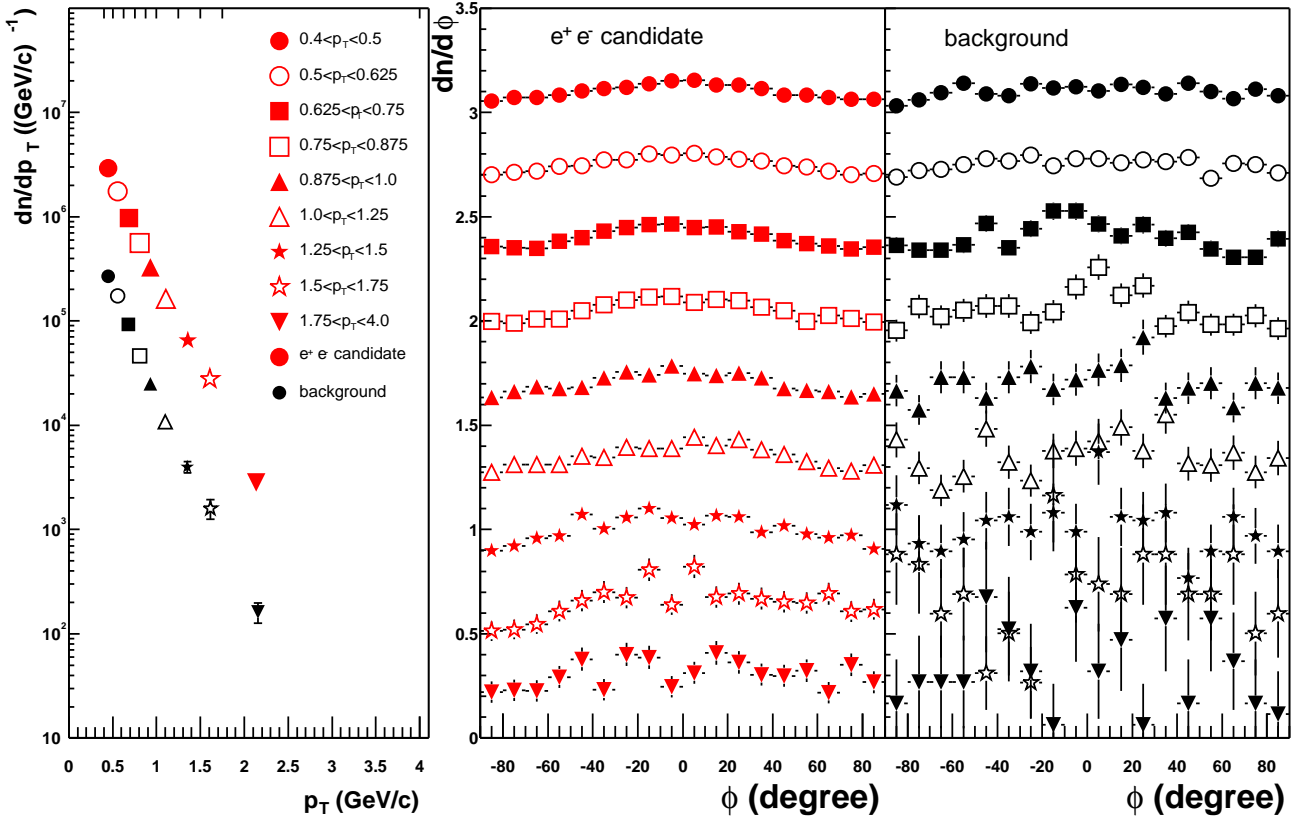}
\caption{\label{fig:azm_ani}
Azimuthal distributions relative to the reaction plane of electrons (middle panel) and the background (right panel)
in each momentum bin are shown. The distributions are overlaid by shifting them on the vertical axis such that the spacing between each is equal.
The raw yields of each distribution are shown the left panel.}
\end{center}
\end{figure*}

The azimuthal distributions of electrons relative to the reaction plane are shown in the middle panel of Fig.~\ref{fig:azm_ani}.
The distributions are overlaid by shifting them on the vertical axis such that the spacing between each is equal.
Each symbol represents the measured $p_{T}$ region indicated in the left panel, which shows
the raw yields of the each distribution with large symbols.
As described in section \ref{sec-chpart}, less than 10\% background remains from accidental RICH associations.
The azimuthal distributions of the background are shown in the right panel, and the yields
are shown as small symbols in the left panel.
The electron $v_{2}$ are measured after subtraction of this background ($dN^{e_{back}}/d\phi$) from electrons 
that are identified by the RICH ($dN^{e_{cand}}/d\phi$):
\begin{equation}
\frac{dN^{e}}{d\phi} = \frac{dN^{e_{cand}}}{d\phi}-\frac{dN^{e_ {back}}}{d\phi}.
\label{eq:EQ11}
\end{equation}

\begin{figure}[htb]
\begin{center}
\includegraphics[width=1.0\linewidth,clip]{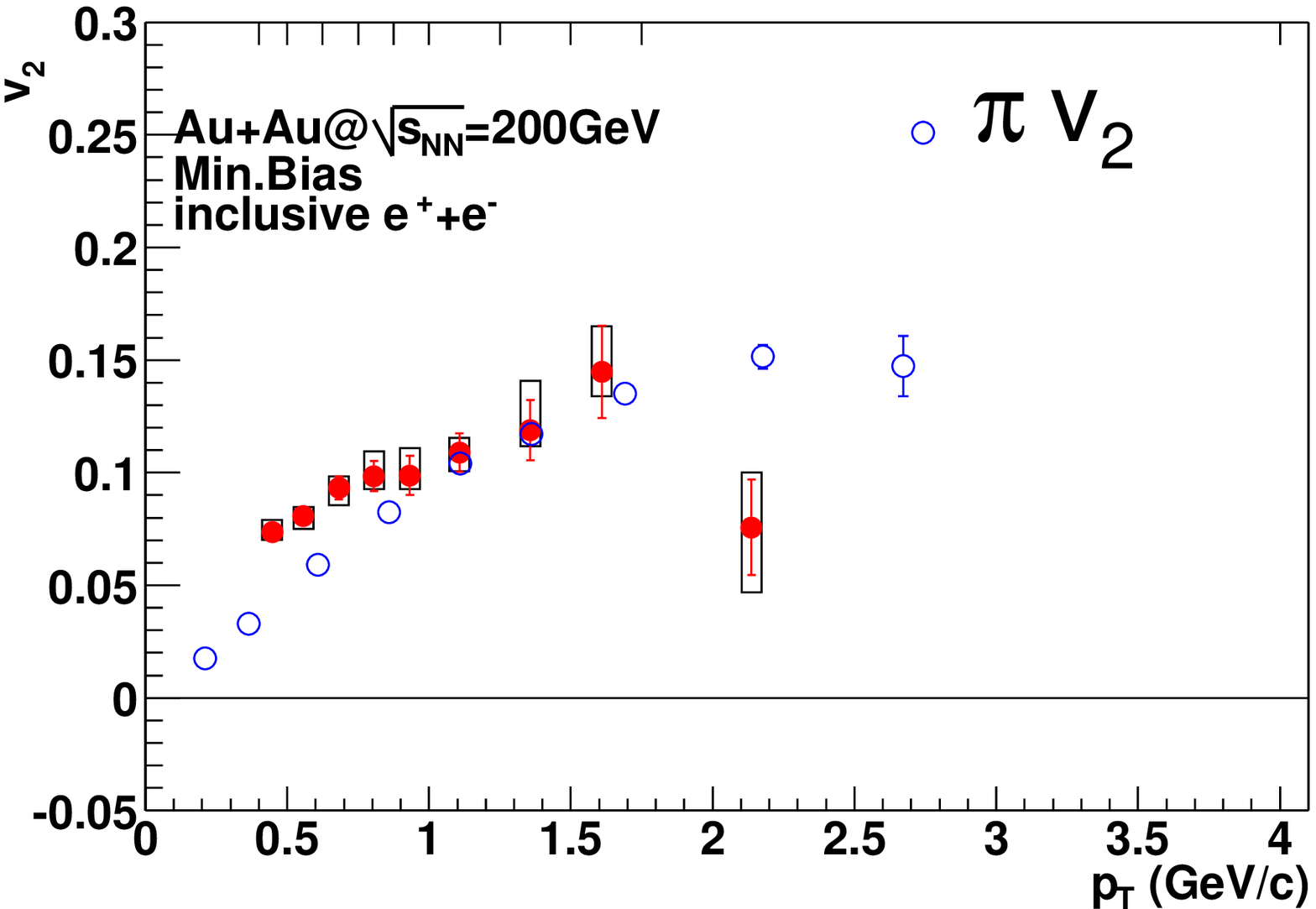}
\caption{\label{fig:pt_ev2}
Transverse momentum dependence of the electron $v_{2}$. 
The upper horizontal scale shows the bin size in $p_{T}$.
The statistical errors are shown as vertical lines in the figure.
The systematic uncertainty from the determination of the reaction plane and 
electron identification are shown as boxes. 
A comparison with $\pi$ $v_{2}$ is also shown.}
\end{center}
\end{figure}

The transverse momentum dependence of $v_{2}$ for electrons after subtracting background is shown
in Fig.~\ref{fig:pt_ev2}.
The statistical errors are shown as vertical lines in the figure.
The 1 $\sigma$ systematic uncertainties are shown as vertical bands.
The systematic uncertainties include the systematic uncertainty of the reaction plane
determination and electron identification.
The systematic uncertainty of the reaction plane determination is about $5 \%$. 
The uncertainty was estimated by measuring $v_{2}$ with reaction plane which was determined by the
North side BBC, the South side BBC, and a combination of the North and South sides.
The systematic uncertainty from electron identification was estimated by measuring electron $v_{2}$ with 
several different sets of electron identification cuts.
A comparison with $v_{2}$ for charged pion \cite{pidflowPHENIX} is also shown in Fig.~\ref{fig:pt_ev2}.
At low $p_{T}$ ($p_{T}<$ 1.0 GeV/$c$), the electron $v_{2}$ is larger than the $v_{2}$ of the pion.
In this region electrons come mainly from $\pi^0$ decays, directly from the Dalitz decays or 
indirectly from photon conversions. 
Due to the fact that the decay angle of the $\pi^{0}$ decay is small, the electron has about the same 
azimuthal angle as the parent $\pi^{0}$, while at the same time the electron $p_{T}$ is smaller 
than the $\pi^{0}$ $p_{T}$.
Therefore the electron $v_{2}$ at a given $p_{T}$ corresponds to the larger $v_{2}$ of the $\pi^{0}$ at higher $p_{T}$.
The $v_{2}$ of charged pions is consistent with that of neutral pions \cite{pi0v2},
therefore the inclusive electron $v_{2}$ is higher than the $\pi$ $v_{2}$.

\begin{figure}[tbh]
\includegraphics[width=1.0\linewidth,clip]{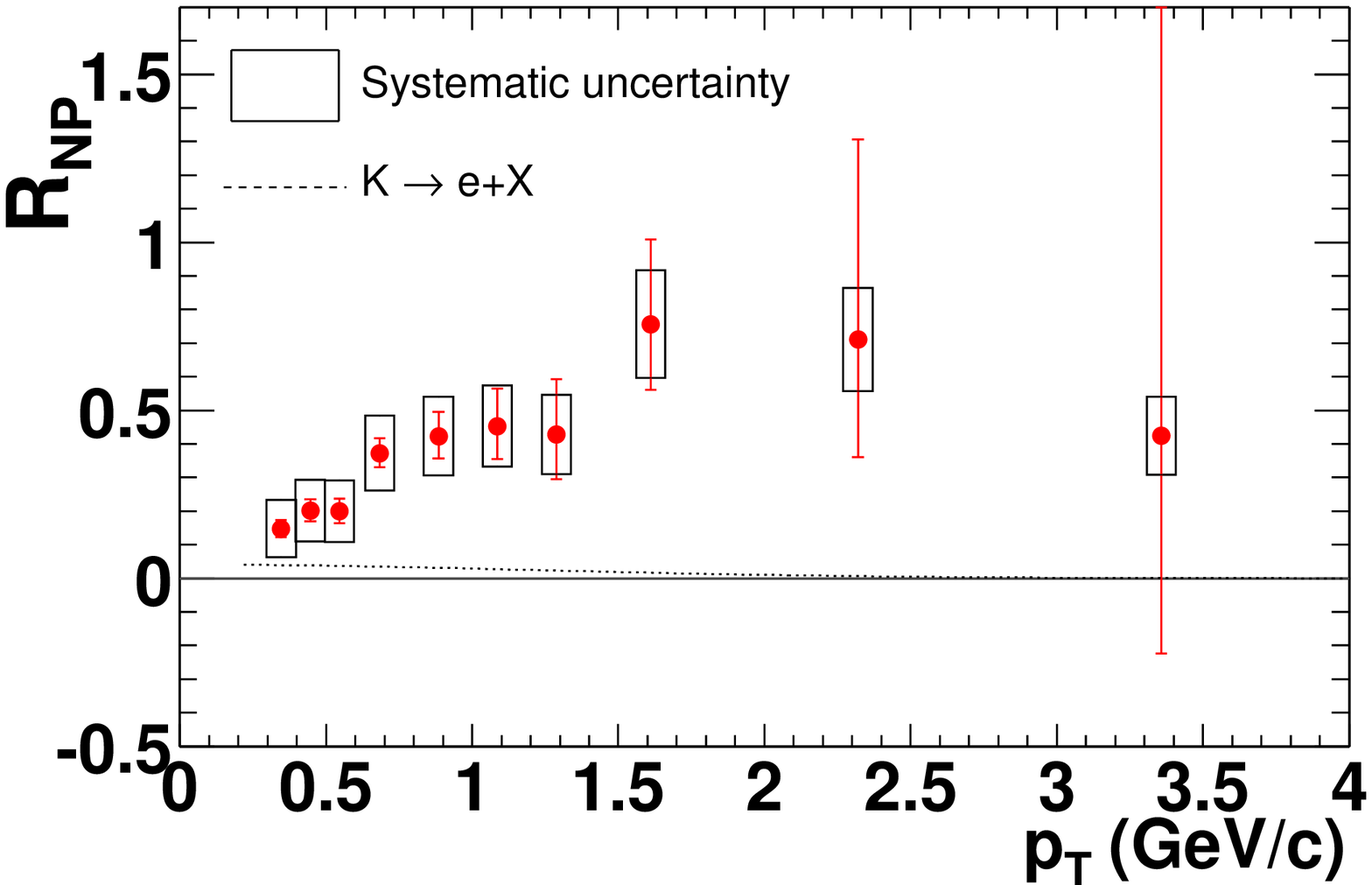}
\caption{\label{fig:npratio}
Ratio of non-photonic to photonic $e^{\pm}$ yields ($R_{NP}$, points) and
contribution from kaon decays (dashed line) \cite{esingle200}.}
\end{figure}

\subsection{Heavy flavor electron $v_{2}$}

The inclusive electron sample has two components: (1)``non-photonic'' - primarily semi-leptonic decays of mesons
containing heavy (charm and bottom) quarks, and (2)``photonic'' - Dalitz decays of light neutral mesons 
($\pi_{0}$, $\eta$, $\eta^{\prime}$, $\omega$ and $\phi$)
and photon conversions in the detector material ~\cite{esingle200}.
The azimuthal distribution of electrons ($dN^{e}/d\phi$) is the sum of the azimuthal distributions of photonic 
electrons ($dN^{\gamma}/d\phi$) and non-photonic electrons ($dN^{non-\gamma}/d\phi$):
\begin{eqnarray}
\frac{dN_{e}}{d\phi}&=&\frac{dN_{e}^{\gamma}}{d\phi}+\frac{dN_{e}^{non-\gamma}}{d\phi}.
\label{eq:EQ12}
\end{eqnarray}
The second harmonic of the Fourier expansion of 
each azimuthal distribution is defined according to
\begin{widetext}
\begin{eqnarray}
N_{e}\left(1+2v_{2_{e}}\cos(2\phi)\right) 
&=&N_{e}^{\gamma}\left(1+2v_{2_{e}}^{\gamma}\cos(2\phi)\right)+N_{e}^{non-\gamma}\left(1+2v_{2_{e}}^{non-\gamma}\cos(2\phi)\right) \nonumber \\
&=&(N_{e}^{\gamma}+N_{e}^{non-\gamma})\left(1+2\frac{N_{e}^{\gamma}v_{2_{e}}^{\gamma}+N_{e}^{non-\gamma}v_{2_{e}}^{non-\gamma}}{N_{e}^{\gamma}+N_{e}^{non-\gamma}}\cos(2\phi)\right),
\label{eq:EQ13}
\end{eqnarray}
\end{widetext}
where $v_{2_{e}}$ is the $v_{2}$ of inclusive electron, $v_{2_{e}}^{\gamma}$ is the $v_2$ of the photonic electrons and $v_{2_{e}}^{non-\gamma}$ 
is the $v_2$ of the non-photonic electrons.
From Eq.~\ref{eq:EQ13}, the inclusive electron $v_{2}$ is given by:
\begin{eqnarray}
v_{2_{e}}&=& \frac{N_{e}^{\gamma}v_{2_{e}}^{\gamma}+N_{e}^{non-\gamma}v_{2_{e}}^{non-\gamma}}{N_{e}^{\gamma}+N_{e}^{non-\gamma}} \nonumber \\
        &=& \frac{N_{e}^{\gamma}v_{2_{e}}^{\gamma}+(N_{e}-N_{e}^{\gamma})v_{2_{e}}^{non-\gamma}}{N_{e}} \nonumber \\
        &=& rv_{2_{e}}^{\gamma}+(1-r)v_{2_{e}}^{non-\gamma},        
\label{eq:EQ14}        
\end{eqnarray}
where $r$ is defined as $r=1/(1+R_{NP})$.
$R_{NP}$ is the ratio of the number of non-photonic electrons to photonic electrons ($N_{e}^{non-\gamma}/N_{e}^{\gamma}$).
We experimentally determined the ratio from analysis of special runs in which additional photon converter was installed.
The details of the method are described in \cite{esingle200}, and the measured ratio is shown in Fig.~\ref{fig:npratio}.
The increase in the number of non-photonic electrons is consistent
with that expected from semileptonic charm decays \cite{esingle200}.
From Eq.~\ref{eq:EQ14} $v_{2_{e}}^{non-\gamma}$ can be expressed as
\begin{equation}
v_{2_{e}}^{non-\gamma} = \frac{v_{2_{e}}-rv_{2_{e}}^{\gamma}}{1-r}.
\label{eq:EQ15}
\end{equation}

\begin{figure*}[bth]
\includegraphics[width=1.0\linewidth,keepaspectratio,clip]{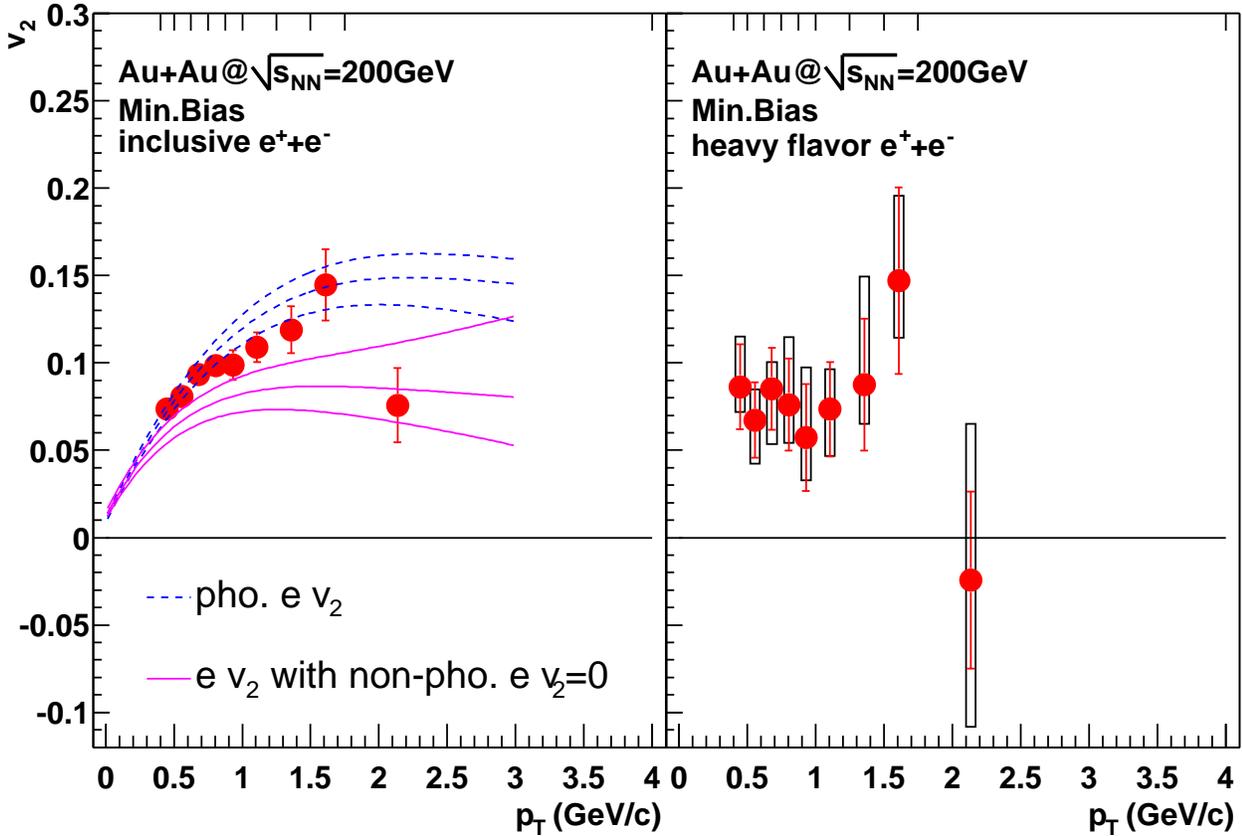}
\caption{\label{fig:rv2}
The left panel : A comparison of the inclusive electron $v_{2}$ with the photonic electron $v_{2}$ (dashed line).
The solid line is $rv_{2_{e}^{\gamma}}$, the photonic electron $v_{2}$
scaled by the ratio of the number of inclusive to photonic electrons. 
The electron $v_{2}$ is the same as $rv_{2_{e}^{\gamma}}$ if the non-photonic electron $v_{2}$ is zero, i.e. the $v_{2}$ of the parent 
particle, such as a $D$ meson, is zero.
The right panel : Transverse momentum dependence of the heavy quark electron $v_{2}$.
The vertical line is the statistical error, which is propagated from 
the statistical errors of the electron $v_{2}$.
The systematic uncertainty from the electron $v_{2}$, the photonic electron $v_{2}$ 
and the ratio $R_{NP}$ is shown as a band.}
\end{figure*}

The dominant sources of photonic electrons are photon conversions and Dalitz decays from $\pi^{0}$ ~\cite{single}.
In addition, we also took into account electrons from $\eta$ decays when calculating photonic electron $v_{2}$.
We assumed that the contributions from $\eta$ decays is 17 $\%$ taken into account $\eta/\pi_{0}$ = 0.45 ~\cite{esingle200}. 
The other sources are ignored when calculating the photonic electron $v_{2}$ due to their small contribution.
The decay electron $v_{2}$ from decay electrons of $\pi^{0}$ and $\eta$ were calculated by Monte Carlo simulation.
The transverse momentum dependence of the $\pi^{0}$ $v_{2}$ was obtained from the
measured $\pi^{0}$ $v_{2}$ \cite{pi0v2} ($p_{T}>$ 1.0 GeV/$c$) and the measured charged $\pi$ $v_{2}$ ($p_{T}<$ 1.0 GeV/$c$).
Both measurements were used since the 
$\pi^{0}$ $v_{2}$ has been measured only above 1.0 GeV/$c$, and both $v_{2}$ measurements are 
consistent at intermediate $p_{T}$ (1.0 $<p_{T}<$ 3.0).  
The measured $\pi^{0}$ spectra \cite{pi0spec} were used to give the input transverse momentum spectrum.
The transverse momentum dependence of the $\eta$ $v_{2}$ was taken to be the same as for the
kaon $v_{2}$.
The transverse momentum spectrum of $\eta$ was approximated by assuming $m_{T}$ scaling of $\pi^{0}$ spectra.
The photonic electron $v_{2}$ that is calculated from the results is shown as the dashed line in the left panel of Fig.~\ref{fig:rv2}.
The middle dashed line is the mean value of the photonic electron $v_{2}$ and the upper and lower 
dashed lines show the 1 $\sigma$ systematic uncertainty.
The systematic uncertainty of the photonic electron $v_{2}$ was estimated from the statistical error and the
systematic error of the measured parent $v_{2}$.
If the non-photonic electron $v_{2}$ is zero, that means the $v_{2}$ of the parent 
particle, such as a $D$ meson, is zero.
Additionally, the inclusive electron $v_{2}$ is the same as that of the scaled photonic electron ($rv_{2_{e}^{\gamma}}$) from Eq.~\ref{eq:EQ14}.
The scaled photonic electron $v_{2}$ is shown as the solid line in the left panel in Fig.~\ref{fig:rv2}.
At intermediate $p_{T}$ ($1.0 <p_{T}< 1.5$) the electron $v_{2}$ is higher than $rv_{2_{e}^{\gamma}}$.
This might suggest that the non-photonic electron has non-zero $v_{2}$ at intermediate $p_{T}$.
The details of this discussion are presented in the next section.

\begin{figure*}[tbh]
\includegraphics[width=0.7\linewidth,keepaspectratio,clip]{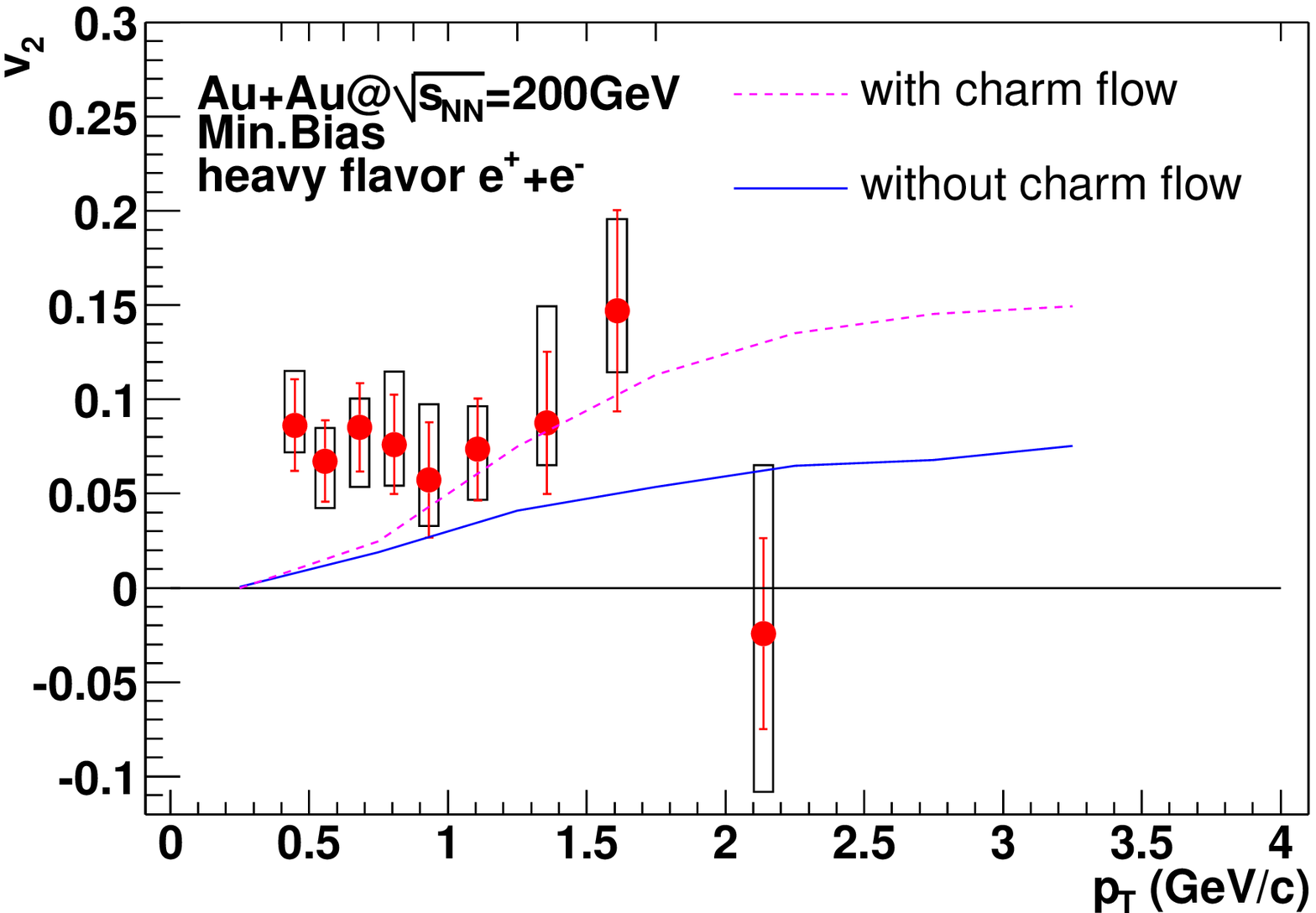}
\caption{\label{fig:h_q_e_v2}
Comparison of the heavy flavor electron $v_{2}$ with two different charm 
flow scenarios from \cite{rapD}. The solid line corresponds to no rescattering of the initially 
produced charm quarks (without flow), while the dashed line reflects the effect of complete thermalization
(with flow).
}
\end{figure*}

Background from kaon decays ($K$ $\to$ $\pi$ $e$ $\nu$) remain in the non-photonic yield.
The contribution of kaon decays to the non-photonic yield, shown in Fig.~\ref{fig:npratio} as dashed line,
is 18 $\%$ at $p_{T} = $ 0.4 GeV/$c$ and decreases rapidly to less than 6 $\%$ for $p_{T} = $ 1 GeV/$c$ \cite{esingle200}.
The transverse momentum dependence of the kaon $v_{2}$ has been measured up to 3.0 GeV/$c$ and that of the
$K^{0}_{S}$ $v_{2}$ has been measured up to 6.0 GeV/$c$ \cite{flowSTAR2}.
The kaon and $K^{0}_{S}$ $v_{2}$ are consistent up to 3.0 GeV/$c$, and the quark coalescence model predicts that these two meson $v_{2}$ values are the same.  
Therefore, kaon and  $K^{0}_{S}$ $v_{2}$ were combined as input for the kaon $v_{2}$.
The transverse momentum spectrum of kaons was obtained from measured kaon spectra up to 2.0 GeV/$c$.
In the high $p_{T}$ region we used scaled $\pi^{0}$ spectra and assumed that the shapes of the kaon spectra were the same as the $\pi^{0}$ spectra,
which are matched with measured kaon spectra around 2.0 GeV/$c$.  

\begin{table*}[tbh]
\caption{\label{tb:sys}
The relative systematic uncertainty of heavy quark electron $v_{2}$.
All errors are given in percent.}
\begin{ruledtabular} \begin{tabular}{cccccccc} 
%                     &    &inc. e $v_{2}$ & pho.e $v_{2}$ & $R_{NP}$   & R.P.  & Total \\ \hline
%Type                 &    &A              & A             & A          & B     & \\ \hline
%0.4 $<p_{T}<$ 1.0    & dw & $<$ 0.40       & $<$ 0.37      & $<$ 0.3   & 0.05  & $<$ 0.58 \\
%                     & up & $<$ 0.73      & $<$ 0.30       & $<$ 0.30  & 0.05  & $<$ 0.80 \\
%1.0 $<p_{T}<$ 1.75   & dw & $<$ 0.27      & $<$ 0.27       & $<$ 0.18  & 0.05  & $<$ 0.43 \\
%                     & up & $<$ 0.70       & $<$ 0.23      & $<$ 0.20  & 0.05  & $<$ 0.75 \\
%1.75 $<p_{T}<$ 4.0   & dw & 2.5           & 0.82           & 1.7       & 0.05  & 3.2 \\
%                     & up & 2.0           & 0.70           & 2.6       & 0.05  & 3.4 \\

$p_{T}$ range        & Sys. error bound  &inclusive e $v_{2}$ ($\%$)& photonic e $v_{2}$ ($\%$) & $R_{NP}$ ($\%$)   & R.P. ($\%$)  & Total ($\%$) \\ \hline
0.4 $<p_{T}<$ 1.0    & lower             & $<$ 32                   & $<$ 26                    & $<$ 21            & 4.5          & $<$ 42 \\
                     & upper             & $<$ 63                   & $<$ 21                    & $<$ 21            & 4.5          & $<$ 70 \\
1.0 $<p_{T}<$ 1.75   & lower             & $<$ 25                   & $<$ 21                    & $<$ 14            & 4.5          & $<$ 36 \\
                     & upper             & $<$ 67                   & $<$ 17                    & $<$ 15            & 4.5          & $<$ 70 \\
1.75 $<p_{T}<$ 4.0   & lower             & 280                      & 78                        & 190               & 4.5          & 340 \\
                     & upper             & 220                      & 64                        & 280               & 4.5          & 360 \\ \hline
Type                 &                   &A                         & A                         & A                 & B            & \\ 
\end{tabular} \end{ruledtabular} \end{table*}

The main source of the non-photonic electrons is semileptonic decays of heavy flavor (charm and beauty). 
Therefore the non-photonic electron $v_{2}$ that was obtained by subtracting photonic electron and kaon decays from inclusive electrons
should be heavy flavor electron \cite{esingle200} $v_{2}$, which reflects the azimuthal anisotropy of heavy quarks.
The result of the heavy flavor electron $v_{2}$ is shown in the right panel of Fig.~\ref{fig:rv2}.
The vertical lines are the statistical errors that are propagated from 
the statistical errors of the inclusive electron $v_{2}$ shown in Fig.~\ref{fig:pt_ev2}.
The $1 \sigma$ systematic uncertainty of heavy flavor electron $v_{2}$ is shown as bands.
The systematic uncertainty includes the systematic uncertainty of the reaction plane, the measured inclusive electron $v_{2}$ (w/o R.P.), 
the photonic electron $v_{2}$ (w/o R.P.) and $R_{NP}$.
The systematic uncertainty of $R_{NP}$ is the quadratic sum of the statistical and systematic errors
because $R_{NP}$ is measured with different data set.
There are two categories of uncertainty: Type A is a point-to-point error uncorrected between $p_{T}$ bins, and
type B is a common displacement of all points by the same factor independent of $p_{T}$. 
The total systematic uncertainty is calculated by propagating the errors on the individual
quantities that enter into Eq.~\ref{eq:EQ15}.
Table \ref{tb:sys} shows the relative systematic uncertainty of heavy quark electron $v_{2}$.

From the result we calculated the confidence level for a non-zero $v_{2}$. 
We assumed that the data of measured heavy flavor electron $v_{2}$ follows a Gaussian distribution and 
the $\sigma$ was obtained by calculating quadratic sum of the statistical and systematic errors of the heavy flavor electron $v_{2}$
assumed these errors are independent.
In the intermediate $p_{T}$ region (1.0 GeV/$c$ $<p_{T}<$ 1.75 GeV/$c$), the confidence level is  90 $\%$,
suggesting that the measured heavy flavor electron $v_{2}$ has a non-zero $v_{2}$. 

Assuming the quark coalescence model,
decay electron $v_{2}$ from $D$ mesons has been predicted~\cite{rapD}.
In the model $D$ mesons are formed from charm quark coalescence with thermal light quarks at hadronization. 
For charm quark momentum spectra, two extreme scenarios are considered. 
The first scenario assumes no reinteractions after the production of charm-anticharm quark 
pairs in initial state hard processes (calculated from PYTHIA). 
The second scenario assumes complete thermalization with the transverse flow of the bulk matter.
%Fig.~\ref{fig:h_q_e_v2} shows the $v_2$ of decay electrons from $D$ mesons in the 
%``no reinteraction'' scenario as a solid line, while the dashed line reflects the ``thermalization'' scenario
%with the measured heavy-flavor electron $v_{2}$ values as presented in this paper.
Figure~\ref{fig:h_q_e_v2} shows a comparison of the heavy flavor electron $v_{2}$ with decay electrons from $D$ mesons in the 
``no reinteraction'' scenario as a solid line, while the dashed line reflects the ``thermalization'' scenario.
Due to large systematic and statistical uncertainty of the current measurement,
neither scenario is excluded by this single electron $v_{2}$ measurement.

\section{Summary}
In summary, we have measured the elliptic flow, $v_{2}$, of single electrons from heavy flavor decay.
This single electron $v_{2}$ is produced by subtracting the $v_{2}$ of electron sources 
such as photon conversion from the $v_{2}$ of inclusive electrons measured with the PHENIX detector 
in Au+Au collisions at $\sqrt{s_{NN}}$ = 200 GeV with respect to the reaction plane 
defined at high rapidities ($|\eta|=3-4$).
The measured heavy flavor electron $v_{2}$ is nonzero with a 90 $\%$ confidence level.
Two model calculations from \cite{rapD} assume extremely different scenarios:
either no reinteraction of the initially produced charm quarks or complete thermalization with
the bulk matter.  Both of these calculations are consistent within errors with the measured heavy flavor electron $v_{2}$.

High luminosity Au+Au collisions at $\sqrt{s_{NN}}$ = 200 GeV have been recorded by the PHENIX 
experiment during Run4 (2003-2004). 
The much higher statistical precision of these data should allow an unambiguous result on the
important issue of charm flow.

\section{Acknowledgements}   % Run-2 long from for PRC, PLB, etc.

We thank the staff of the Collider-Accelerator and Physics
Departments at Brookhaven National Laboratory and the staff
of the other PHENIX participating institutions for their
vital contributions.  We acknowledge support from the
Department of Energy, Office of Science, Nuclear Physics
Division, the National Science Foundation, Abilene Christian
University Research Council, Research Foundation of SUNY, and
Dean of the College of Arts and Sciences, Vanderbilt
University (U.S.A), Ministry of Education, Culture, Sports,
Science, and Technology and the Japan Society for the
Promotion of Science (Japan), Conselho Nacional de
Desenvolvimento Cient\'{\i}fico e Tecnol{\'o}gico and Funda\c
c{\~a}o de Amparo {\`a} Pesquisa do Estado de S{\~a}o Paulo
(Brazil), Natural Science Foundation of China (People's
Republic of China), Centre National de la Recherche
Scientifique, Commissariat {\`a} l'{\'E}nergie Atomique,
Institut National de Physique Nucl{\'e}aire et de Physique
des Particules, and Institut National de Physique
Nucl{\'e}aire et de Physique des Particules, (France),
Bundesministerium f\"ur Bildung und Forschung, Deutscher
Akademischer Austausch Dienst, and Alexander von Humboldt
Stiftung (Germany), Hungarian National Science Fund, OTKA
(Hungary), Department of Atomic Energy and Department of
Science and Technology (India), Israel Science Foundation
(Israel), Korea Research Foundation and Center for High
Energy Physics (Korea), Russian Ministry of Industry, Science
and Tekhnologies, Russian Academy of Science, Russian
Ministry of Atomic Energy (Russia), VR and the Wallenberg
Foundation (Sweden), the U.S. Civilian Research and
Development Foundation for the Independent States of the
Former Soviet Union, the US-Hungarian NSF-OTKA-MTA, the
US-Israel Binational Science Foundation, and the 5th European
Union TMR Marie-Curie Programme.

%\clearpage

\end{document}